\documentclass[a4paper]{article}

 \usepackage[frenchb,english]{babel}

\usepackage[dvips]{graphicx}
\usepackage{amsmath}
\usepackage{epsfig}
\usepackage[latin1]{inputenc}
\usepackage[left]{lineno}

\usepackage{amstext}
\usepackage{amsmath}
\usepackage{amsbsy}
\usepackage{multirow}
\usepackage{amssymb}
\usepackage{amsfonts}
\usepackage{wasysym}
\usepackage{epsfig}
\usepackage{subfigure}
 \usepackage[left]{lineno}
\usepackage{hyperref}
\newcommand{\ANKARA}      {4}
\newcommand{\ANNECY}      {12}
\newcommand{\BARI}        {35}
\newcommand{\BARIINFN}    {21}
\newcommand{\BERN}        {7}
\newcommand{\BOLOGNA}     {15}
\newcommand{\BOLOGNAINFN} {16}
\newcommand{\BRUSSELS}    {38}

\newcommand{\DUBNA}      {18}
\newcommand{\FRASCATI}   {17}
\newcommand{\FUNABASHI}  {27}	
\newcommand{\HAIFA}      {28}
\newcommand{\HAMBURG}    {24}
\newcommand{\GAZWADONG}  {32}
\newcommand{\KARIYA}     {33}
\newcommand{\KOBE}       {6}
\newcommand{\LAQUILA}    {22}
\newcommand{\LNGS}       {19}
\newcommand{\LYON}       {8}
\newcommand{\MOSCOWINR}  {1}
\newcommand{\MOSCOWITEP} {25}
\newcommand{\MOSCOWLPI}  {3}
\newcommand{\MOSCOWSINP} {5}
\newcommand{\MUNSTER}    {26}
\newcommand{\NAGOYA}     {30}
\newcommand{\NAPOLI}     {20}
\newcommand{\NAPOLIINFN} {2}
\newcommand{\PADOVA}     {14}
\newcommand{\PADOVAINFN} {11}
\newcommand{\ROMA}       {34}
\newcommand{\ROSTOCK}    {37}
\newcommand{\SALERNO}    {13}
\newcommand{\STRASBOURG} {23}
\newcommand{\TUNIS}      {10}
\newcommand{\URBINO}     {29}
\newcommand{\UTSUNOMIYA} {36}
\newcommand{\ZAGREB}     {31}
\newcommand{\ZURICH}     {9}
\newcommand{\CORR}       {*}
\newcommand{\NOWLLR}	 {a}
\newcommand{\NOWLNGS}     {b}
\newcommand{\NOWINAFIASF} {c}
\newcommand{\NOWROMA}     {d}
\newcommand{\NOWPUSAN}    {e}
\newcommand{\NOWFRASCATI} {f}
\newcommand{\NOWASAN}     {g}
\newcommand{\NOWSUBATECH} {h}
\newcommand{\NOWBERN}     {i}

\newcommand{\OperaInstitutes}{
\MOSCOWINR   . INR-Institute for Nuclear Research of the Russian Academy of Sciences, RUS-117312 Moscow, Russia \\
\NAPOLIINFN  . INFN Sezione di Napoli, I-80125 Napoli, Italy \\
\MOSCOWLPI   . LPI-Lebedev Physical Institute of the Russian Academy of Sciences, 119991 Moscow, Russia \\
\ANKARA      . METU-Middle East Technical University, TR-06531 Ankara, Turkey \\
\MOSCOWSINP  . SINP MSU-Skobeltsyn Institute of Nuclear Physics of Moscow State University, RUS-119992 Moscow, Russia \\
\KOBE        . Kobe University, J-657-8501 Kobe, Japan \\
\BERN        . Albert Einstein Center for Fundamental Physics, Laboratory for High Energy Physics (LHEP),
               University of Bern, CH-3012 Bern, Switzerland \\
\LYON        . IPNL, Universit\'e Claude Bernard Lyon 1, CNRS/IN2P3, F-69622 Villeurbanne, France \\
\ZURICH      . ETH Zurich, Institute for Particle Physics, CH-8093 Zurich, Switzerland \\
\TUNIS       . Unit\'e de Physique Nucl\'eaire et des Hautes Energies (UPNHE), Tunis, Tunisia \\
\PADOVAINFN  . INFN Sezione di Padova, I-35131 Padova, Italy \\
\ANNECY      . LAPP, Universit\'e de Savoie, CNRS/IN2P3, F-74941 Annecy-le-Vieux, France \\
\SALERNO     . Dipartimento di Fisica dell'Universit\`a  di Salerno and INFN, I-84084 Fisciano, Salerno, Italy \\
\PADOVA      . Dipartimento di Fisica dell'Universit\`a  di Padova, I-35131 Padova, Italy \\
\BOLOGNA     . Dipartimento di Fisica dell'Universit\`a  di Bologna, I-40127 Bologna, Italy \\
\BOLOGNAINFN . INFN Sezione di Bologna, I-40127 Bologna, Italy \\
\FRASCATI    . INFN - Laboratori Nazionali di Frascati dell'INFN, I-00044 Frascati (Roma), Italy \\
\DUBNA       . JINR-Joint Institute for Nuclear Research, RUS-141980 Dubna, Russia \\
\LNGS        . INFN - Laboratori Nazionali del Gran Sasso, I-67010 Assergi (L'Aquila), Italy \\
\NAPOLI      . Dipartimento di Scienze Fisiche dell'Universit\`a Federico II di Napoli, I-80125 Napoli, Italy \\
\BARIINFN    . INFN Sezione di Bari, I-70126 Bari, Italy \\
\LAQUILA     . Dipartimento di Fisica dell'Universit\`a dell'Aquila and INFN, I-67100 L'Aquila, Italy \\
\STRASBOURG  . IPHC, Universit\'e de Strasbourg, CNRS/IN2P3, F-67037 Strasbourg, France \\
\HAMBURG     . Hamburg University, D-22761 Hamburg, Germany \\
\MOSCOWITEP  . ITEP-Institute for Theoretical and Experimental Physics, RUS-117259 Moscow, Russia \\
\MUNSTER     . University of M\"unster, D-48149 M\"unster, Germany \\
\FUNABASHI   . Toho University, J-274-8510 Funabashi, Japan \\
\HAIFA       . Department of Physics, Technion, IL-32000 Haifa, Israel \\
\URBINO      . Universit\`a degli Studi di Urbino "Carlo Bo", I-61029 Urbino, Italy \\
\NAGOYA      . Nagoya University, J-464-8602 Nagoya, Japan \\
\ZAGREB      . IRB-Rudjer Boskovic Institute, HR-10002 Zagreb, Croatia \\
\GAZWADONG   . Gyeongsang National University, ROK-900 Gazwa-dong, Jinju 660-300, Korea \\
\KARIYA      . Aichi University of Education, J-448-8542 Kariya (Aichi-Ken), Japan \\
\ROMA        . Dipartimento di Fisica dell'Universit\`a  di Roma ``La Sapienza" and INFN, I-00185 Roma, Italy \\
\BARI        . Dipartimento di Fisica dell'Universit\`a  di Bari, I-70126 Bari, Italy \\
\UTSUNOMIYA  . Utsunomiya University, J-321-8505 Tochigi-Ken, Utsunomiya, Japan \\
\ROSTOCK     . Fachbereich Physik der Universit\"at Rostock, D-18051 Rostock, Germany \\
\BRUSSELS    . IIHE, Universit\'e Libre de Bruxelles, B-1050 Brussels, Belgium \\
\NOWLLR	 . Now at Laboratoire Leprince-Ringuet, CNRS/IN2P3 Ecole polytechnique, F-91128 Palaiseau, France \\
\NOWLNGS     . Now at INFN - Laboratori Nazionali del Gran Sasso, I-67010 Assergi (L'Aquila), Italy \\
\NOWINAFIASF . Now at INAF/IASF, Sezione di Milano, I-20133 Milano, Italy \\
\NOWROMA     . Now at Dipartimento di Fisica dell'Universit\`a  di Roma ``La Sapienza" and INFN, I-00185 Roma, Italy \\
\NOWPUSAN    . Now at Pusan National University, Geumjeong-Gu Busan 609-735, Korea \\
\NOWFRASCATI . Now at INFN - Laboratori Nazionali di Frascati dell'INFN, I-00044 Frascati (Roma), Italy \\
\NOWASAN     . Now at Asan Medical Center, 388-1 Pungnap-2 Dong, Songpa-Gu, Seoul 138-736, Korea \\
\NOWSUBATECH . Now at SUBATECH, CNRS/IN2P3, F-44307 Nantes, France \\
\NOWBERN     . Now at Albert Einstein Center for Fundamental Physics, Laboratory for High Energy Physics (LHEP),
               University of Bern, CH-3012 Bern, Switzerland \\
\CORR          Corresponding Authors: \\
               Email addresses: magali.besnier@llr.in2p3.fr (M. Besnier),
			duchesneau@lapp.in2p3.fr (D. Duchesneau) \\
               }

\newcommand{\OperaAuthorList}{
N.~Agafonova$^{\MOSCOWINR}$, 
A.~Aleksandrov$^{\NAPOLIINFN,\MOSCOWLPI}$, 
O.~Altinok$^{\ANKARA}$, 
A.~Anokhina$^{\MOSCOWSINP}$, 
S.~Aoki$^{\KOBE}$, 
A.~Ariga$^{\BERN}$, 
T.~Ariga$^{\BERN}$,
D.~Autiero$^{\LYON}$, 
A.~Badertscher$^{\ZURICH}$, 
A.~Bagulya$^{\MOSCOWLPI}$
% ...
\\
}

\begin{document}
\pagestyle{empty} 

\title{
{\bf Momentum measurement by the Multiple Coulomb Scattering method 
in the OPERA lead emulsion target}\\
\vspace{1cm}
OPERA COLLABORATION
}

\maketitle
\thispagestyle{empty} 

\author{\noindent \\ \OperaAuthorList }

\begin{flushleft}
\footnotesize{\OperaInstitutes }
\end{flushleft}


\begin{center}
Submitted to \it {New Journal of Physics}
\end{center}

\vspace{0.5cm}

\begin{abstract}
\noindent
A new method of momentum measurement of charged particles through
Multiple†Coulomb Scattering (MCS) in the OPERA lead emulsion target is presented.
It is based on  precise measurements of track angular deviations performed 
thanks to the very high resolution of nuclear emulsions. The algorithm has been tested
with Monte Carlo (MC) pions. The  results are found to describe
within the expected uncertainties the data obtained from test beams. 
†We also report a comparison of muon momenta evaluated through MCS in the OPERA lead emulsion target
 with those determined by the 
electronic detectors for  neutrino charged current interaction events. The two independent measurements
agree within the experimental uncertainties, and the results validate the algorithm
developed for the  emulsion detector  of OPERA.

\end{abstract}

\newpage
\pagestyle{plain}
\setcounter{page}{1}
\setcounter{footnote}{0}

\section{Introduction}
\par $\ $
\par
The momentum of charged particles can be measured in Emulsion Cloud
Chambers (ECC)~\cite{ECC} made of  massive  material plates, used as target, interleaved with
 nuclear emulsion films acting as high resolution tracking devices. This technique was exploited by
the DONUT experiment~\cite{Donut} and is currently used in the OPERA
experiment searching for $\nu_\mu \rightarrow \nu_\tau$ oscillations in the CNGS neutrino beam~\cite{opera}. 
The study shown in this paper uses the geometry and the characteristics of the 
O\-PE\-RA neutrino target ECC elements called  ``bricks". 
They have dimensions of
12.7$\times$10.2$\times$7.5 cm$^3$  and are composed of a sequence of
56  lead plates (1 mm thick) and  57 emulsion films (44 $\mu$m thick emulsion
layers on each side of a 205 $\mu$m thick plastic base). The total length of a brick corresponds to
 about 10$X_0$.\\
Charged particles crossing the emulsions ionise silver bromide crystals, and 
clusters of silver grains, appearing as black dots, are formed along their  
paths after film processing. Automatic microscopes~\cite{scan} are used to reconstruct 
3D particle track segments. 
Micro-track segments are reconstructed in single emulsion layers as sequences of aligned grains.
 Two matching micro-tracks in a film define a base-track, obtained as the straight line connecting 
the grains closest to the plastic base in the two emulsion layers.

A track reconstructed through connecting 
segments in two or more films is called a volume-track.

\par The momentum measurement by Multiple Coulomb Scattering (MCS) can be performed by using either
the track position (coordinate method)~\cite{Kodama:2002dk} or the track angle (angular method)~\cite{Lellis:2003xt} 
measured in each emulsion film. The two methods determine the deviations of the trajectory
from a straight line on the basis of position or angle measurements,
respectively.  The use of one method rather than the other depends on
the required accuracy,  and on the achievable spatial and angular resolutions. 
In OPERA ECC bricks, base-track directions are measured with a precision of a few mrad. 
Moreover, the angular method does not depend on a precise
knowledge of the relative alignment of the different emulsion films.
The evolution of slopes of consecutive base-tracks forming a volume-track can thus be used
to compute the mean Coulomb scattering angle in a given lead thickness,
which is directly related to the particle momentum.
 The angular resolution of the emulsions allows  the determination of
charged particle  momentum  from several hundreds of MeV/c to a few GeV/c, which corresponds to the momentum range of secondary 
hadrons produced in  neutrino interactions in the OPERA experiment. 
Several approaches to perform the  angular deviation measurements 
in lead have been tested and compared in previous studies.
The  method presented in this paper is based on the work detailed in~\cite{thesis} and
is used in the analysis of the neutrino events observed in OPERA~\cite{tau_candidate}.
\par The first part describes the  method and the special treatment used for the large angle
tracks. Results from Monte Carlo and from data  analysis  with pions from 1 to
8 GeV/c momentum for various track lengths are summarised in sections~\ref{sec:MC} and~\ref{sec:data}. 
In the last section the results from the application of the algorithm to
 muon tracks reconstructed independently in the OPERA electronic detectors with momenta below 6 GeV/c are presented.

\section{Measurement method}\label{sec:explain}

\subsection{Scattering angle dependence on the lead thickness}
The main ingredient for the angular method measurement is the
availability of several angular measurements along
a volume-track. The present approach uses
the angle differences measured in pairs of emulsion films separated
by lead. 
In the following, one cell corresponds to one lead plate and one film.
Figure \ref{fig:brick1} provides a schematic view of a volume-track and its associated
base-tracks in the XZ projection plane.
Let $\theta_i$ be the angle of a
given base-track in the $i^{th}$ emulsion film,
$i\in\left\{1,\ldots,57\right\}$ in the XZ or YZ projection plane. Defining
$\theta_{ik}$ = $\theta_{i+k} - \theta_i$ as the scattering
angle after crossing a number k of cells,
its distribution is peaked at zero and has a shape that can be approximated by a
Gaussian with a standard deviation given by \cite{pdg}
\begin{linenomath}
\begin{equation}
\theta_{0}=\frac{13.6}{(pc\beta)}\times\sqrt{\frac{x}{X_{0}}}\times\left[1+0.038\ln\left(\frac{x}{X_0}\right)\right]
\label{eq1}
\end{equation}
\end{linenomath}
where $p$ is the particle momentum in MeV/c, $\beta c$ its velocity, $x$ is the
distance traversed and $X_{0}$ is the
radiation length in the  material. The accuracy of this
approximation of  Moliere's theory of scattering is better
than 11\% in any material, with $0.001 < x/X_{0} < 100$~\cite{mcsform} for single charged particles with $\beta \approx$  1.\\
The scattering is dominated by the lead since the radiation length
in the emulsion layers and the plastic base is larger by more than one order
of magnitude. 
For this reason, the value $X_0 = 5.6$  mm will be assumed in the analysis and a thickness of 1mm will be used for each cell, 
neglecting the emulsion films.
By denoting the number of cells
crossed by a particle track as $N_{cell}$, the above-mentioned expression becomes: \begin{linenomath}
\begin{equation}\label{eq3}
\theta_{0}\approx\frac{13.6}{(pc\beta)}\times\sqrt{\frac{N_{cell}}{5.6}}\times\left[1+0.038\ln\left(\frac{N_{cell}}{5.6}\right)\right].
\end{equation}
\end{linenomath}
 The variance of the scattering angle distribution
for a given cell depth $N_{cell}$ = $k$ is given by
\begin{linenomath}
\begin{equation}
    \left\langle \theta^{2}_{meas}\right\rangle_k = \sum_i (\theta_{ik})^2/N_{meas}= \theta^{2}_{0} + \delta\theta^{2},
\end{equation}
\end{linenomath}
where $N_{meas}$ is the number of scattering angle measurements and $\delta\theta$ is an additional term corresponding to the 
base-track angular resolution~\footnote{$\delta \theta$
is the angular resolution between 2 base-tracks. The
 single base-track angular resolution is 
$\delta\theta_{s}$ = $\frac{\delta \theta}{\sqrt{2}}$.
}. \\

\begin{linenomath}
\begin{figure}[hbt]
\begin{center}
\includegraphics[width=150mm]{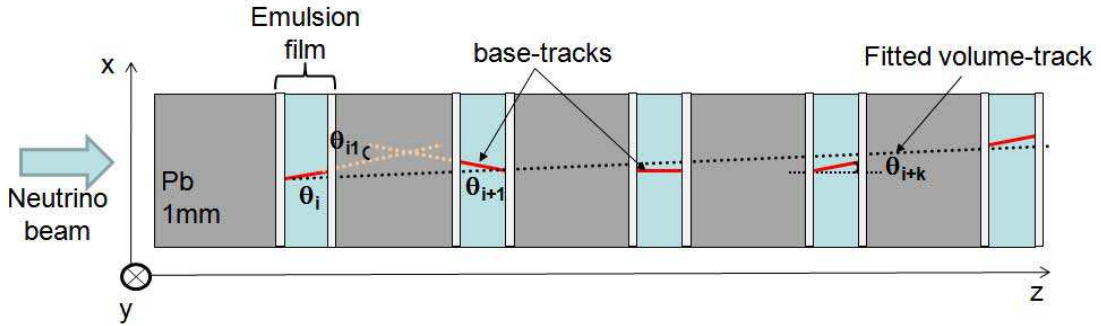}
\vspace*{-0.2cm} \caption{\small{Sketch of 5 lead cells in a  target
brick, where a volume-track and its base tracks are represented in the XZ projection.}}
\label{fig:brick1}
\end{center}
\end{figure}
\end{linenomath}
The current experimental value of $\delta \theta$ is about 2.1 mrad. In order to determine $p$ up to a few GeV/c through the scattering angle, a fit 
of the dependence of $\theta_{meas}$ on the number of crossed cells is performed, treating $p$ as a free parameter and fixing the angular resolution. 
With increasing $p$, the MCS starts dominating over $\delta \theta$  at larger values of $N_{cell}$, where the number of available measurements decreases, thus increasing the statistical error. 
In order to
improve the sensitivity to high-momentum tracks, it is important to
reduce the statistical uncertainty at large crossed thicknesses.
\par The method is illustrated in Figure~\ref{fig:brick2}.
 It consists of using the differences between 
all combinations of pairs of angles separated by $N_{cell}$ cells. For a given cell depth $N_{cell}$ 
and a total track span $N_{pl}$  measured 
as the number of lead plates traversed by the particle, the number of available measurements $N_{meas}$ 
is given by:
\begin{linenomath}
\begin{equation}
N_{meas}=\sum_{i=1}^{N_{cell}}int \left [\frac{N_{pl}-i+1}{N_{cell}}\right ]
\end{equation}
\end{linenomath}

\begin{linenomath}
\begin{figure}[hbt]
\begin{center}
\includegraphics[width=150mm]{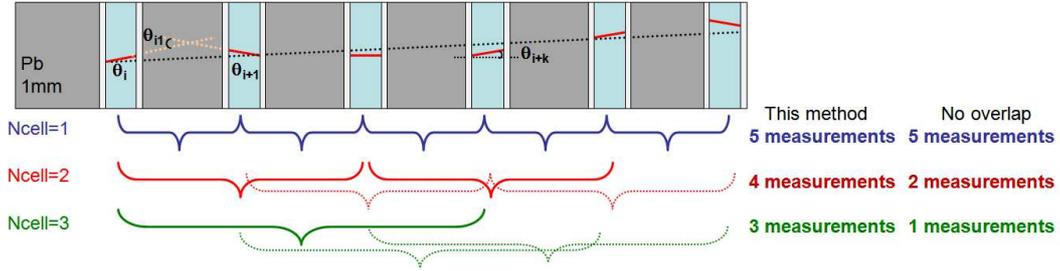}
\vspace*{-0.2cm} \caption{\small{Representation of the number of possible
measurements available when applying the MCS method up to  $N_{cell}$ = 3}. 
} \label{fig:brick2}
\end{center}
\end{figure}
\end{linenomath}
\subsection{Track angle-dependence}

For large-angle tracks the following effects have to be taken into account. 
First, the crossed lead thickness varies as $1/$cos$\theta$,  with $\theta$ being the track angle measured with respect to the
 normal to the emulsion plane (Z coordinate). Second, also the angular resolution $\delta\theta$ depends  on $\theta$, 
as the longitudinal
 uncertainty affects the measured grain  positions along the optical Z axis. This effect is
 dominated by the vertical resolution of the scanning system, and is about 2.5 $\mu$m \cite{scan}.
For angles above 200 mrad, this uncertainty is one order of magnitude larger than that in the transverse X and Y
coordinates.

In order to decouple the intrinsic angular resolution from the slope-dependent contribution, the algorithm is constructed
in a new reference coordinate system. 
It uses  transverse and longitudinal coordinates (denoted respectively as T and L) as defined in~\cite{scan}, 
projected on $\theta_T$ and $\theta_L$ axes of the reference frame schematically shown in Figure \ref{fig:referential}. 
The T and L coordinates are obtained from X and Y by applying a rotation: \begin{linenomath}\begin{equation}\label{eq10a}
\theta_T = \theta_Xcos(\phi)+\theta_Ysin(\phi)
\end{equation}
\begin{equation}\label{eq10b}
\theta_L = -\theta_Xsin(\phi)+\theta_Ycos(\phi)
\end{equation}
where $\phi$ = arctan($\frac{\theta_Y}{\theta_X}$). The 3-dimensional (3D) space angle can be written as
\begin{equation}
tan(\theta_{3D})=\sqrt{tan^2(\theta_X)+tan^2(\theta_Y)}=\sqrt{tan^2(\theta_T)+tan^2(\theta_L)}.
\end{equation}
\end{linenomath}
\begin{linenomath}
\begin{figure}[!h]
\begin{center}
\includegraphics[width=75mm,height=70mm]{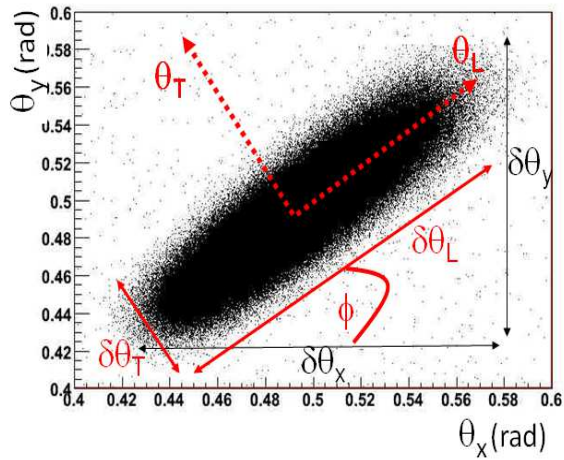}
\vspace*{-0.2cm} \caption{\small{Schematic view of T-L coordinate reference frame, superimposed on
the  $\theta_y$ \emph{versus} $\theta_x$ plot for the base tracks of 10 GeV/c MC  muons at large angle 
($\theta_X=\theta_Y$ = 500 mrad).}} \label{fig:referential}
\end{center}
\end{figure}
\end{linenomath}
\begin{linenomath}
\begin{figure}[!h]
\begin{center}
\includegraphics[width=90mm]{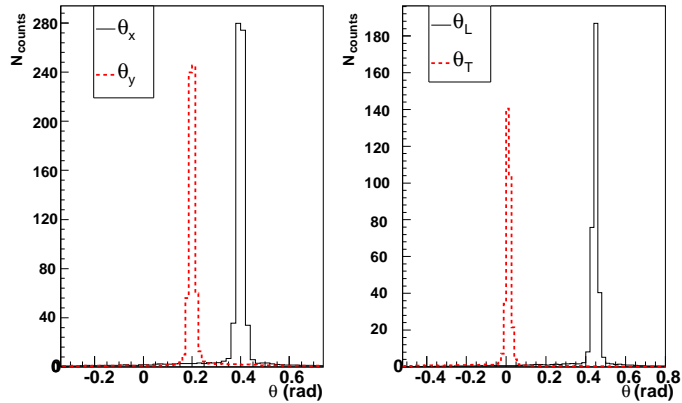}
\vspace*{-0.2cm} \caption{\small{Angular distributions of base tracks from 4 GeV/c MC pions simulated with
 $\theta_X$ = 400 mrad and $\theta_Y$ = 200 mrad in XY (left)  and TL projection planes (right).}} \label{fig:thetaXYTL}
\end{center}
\end{figure}
\end{linenomath}
As can be seen in Figure \ref{fig:referential}, the T coordinate gives an angular spread which remains the same for
 any track angle.  The angular dependence of the resolution can be parameterised as~\cite{alb}:
\begin{linenomath}
  \begin{equation}\label{eq:theta1}
 \delta \theta_T(\theta)=\delta \theta_T(0)= \delta \theta_{3D}(0),
\end{equation}
\end{linenomath}
and
\begin{linenomath}
\begin{equation}\label{eq:theta2}
\delta \theta_L(\theta)=\delta \theta_L(0) +
\epsilon_z tan\theta\ 
\end{equation}
\end{linenomath}
where $\epsilon_z$ is a parameter that linearly depends on the longitudinal uncertainty.

This transformation allows keeping $\theta_T$ centred around 0 mrad as shown in Figure \ref{fig:thetaXYTL}.
As discussed in section \ref{sec:analm}, an unbiased algorithm would use the 3D coordinate (both T and L or X and Y measurements) 
for small angles, and only the T coordinate at large angles. However, the latter choice results in only half the statistics,
even though it is angle-independent and free of bias.
 In the following, all the results are obtained using the T-L coordinate system.

\subsection{Momentum and resolution estimate}
\par In order to estimate the momentum resolution, samples of same-momentum tracks can be analysed. Assuming
a Gaussian distribution for $\theta_0$, the shape of the momentum
distribution can be approximated  by the function
\begin{linenomath}
\begin{equation}\label{eq4}
f(p)=\frac{p_0}{p^2}\times\exp{\left(-\frac{(1/p-1/p_1)^2}{p_2^2}\right)}
\end{equation}
\end{linenomath}
where $p_0$, $p_1$, and $p_2$ are free parameters. The parameter $p_1$ corresponds to the average of the
reconstructed momenta $p_{mean}$. Figure \ref{fig:4gev} (left)
shows an example of this fitted  distribution for
tracks of 4 GeV/c Monte Carlo pions passing through 56 cells.
In order to take into account possible uncertainties coming from the Gaussian analytic approximation of
$1/p$, the mean reconstructed momentum is obtained from the average  
 fits of the distributions of both momentum and inverted momentum (Figure~\ref{fig:4gev} (right)). 
The first one has sensitivity to the high reconstructed 
 momentum tail while the second is more
sensitive to the lower reconstructed momentum values. The difference of the two results is the
systematic uncertainty of the average fitted momentum determination. In the previous example, the results give
 $\langle p \rangle$ = 3.97 $\pm$ 0.01 (stat) $\pm$ 0.08 (syst) GeV/c.

\begin{linenomath}
\begin{figure}[hbt]
\begin{center}
\includegraphics[width=120mm]{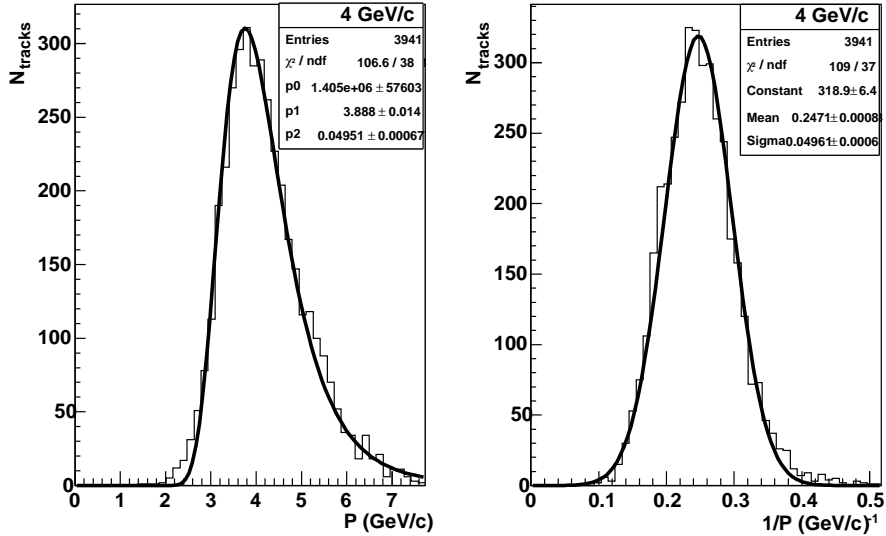}
\vspace*{-0.2cm} \caption{\small{Momentum distribution (left) and inverted momentum distribution (right) for about 4000 tracks 
of 4~GeV/c MC pions reconstructed in an ECC brick.}} \label{fig:4gev}
\end{center}
\end{figure}
\end{linenomath}

\par Since the inverted momentum distribution 1/$p$
has a Gaussian shape, 
 the width of the Gaussian divided by
$1/p_{mean}$  directly gives the momentum resolution estimate $\frac{\Delta(1/p)}{(1/p)}$. 
Its uncertainty can be obtained by propagating the 
errors on the two components which are the width of the distribution and the reconstructed momentum. 
 Therefore, the momentum resolution of the 4 GeV/c MC sample 
of pions passing through an entire OPERA target brick is  20.1 $\pm$ 0.6~\%.

\section{Monte Carlo results} \label{sec:MC}
\par In this section, the results obtained from  Monte Carlo simulations are reported.
The MC data correspond to 2, 4, 6 and 8 GeV/c pion samples of 1000 events each that have been generated
with the simulation tool ORFEO, based on GEANT and developed in the OPERA
framework~\cite{alb}. It simulates particle interactions inside a brick and 
includes the main experimental effects such as the track efficiency and spatial resolutions.
\par This section is divided into two parts: the first  gives the results for small incident angles ($\theta <$ 200 mrad),
 the second for large incident angles ($\theta >$ 200 mrad).

\subsection{Tracks at small incident angles} \label{sec:anasm}
\par Figure \ref{fig:ncellall} shows the dependence of the scattering
angle on $N_{cell}$ for different momenta from 1 to 8 GeV/c~\footnote{The MC samples have been tuned in
order to reproduce the measured $\delta \theta_s$, obtained in the
scanning of the  test beam data samples with same momenta.}. 
Since the MC samples contain a large number of tracks with the same momentum, the single base-track angular
resolution $\delta \theta_s$ can be directly determined together with the particle momentum from the
fits of  Figure~\ref{fig:ncellall}.
Results are summarised in Table \ref{tab:dtMC}.

\begin{linenomath}
\begin{figure}[!h]
\begin{center}
\includegraphics[width=100mm]{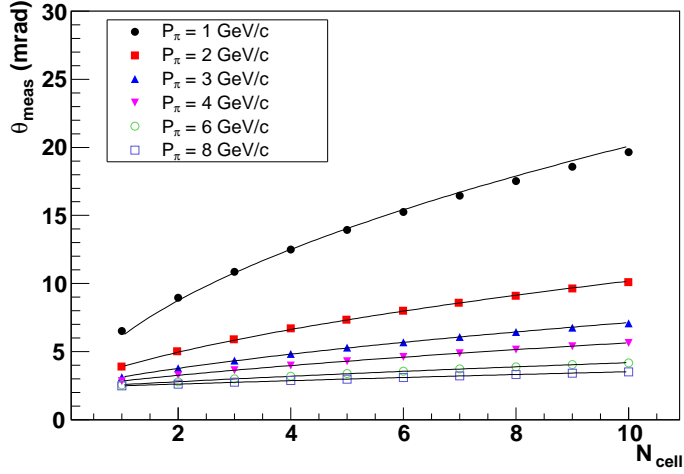}
\vspace*{-0.2cm} \caption{\small{$\theta_{meas}$ dependence on $N_{cell}$ for MC pions of different energies, 
where $\delta \theta_s$ 
has been simulated at a value of $\delta \theta_s^{MC}$ = 1.67 mrad. The solid curves correspond to the  fitted expectations.}}
\label{fig:ncellall}
\end{center}
\end{figure}
\end{linenomath}

\begin{linenomath}
\begin{table}[!h]
 \begin{center}
\begin{tabular}{|c|c|c|c|}

\cline{1-4} $p_{MC}$ (GeV/c) &
 $\delta \theta_s$ (mrad)& $\langle p \rangle$ (GeV/c)&
$\frac{\Delta (1/p)}{(1/p)}(\%$)\\

\cline{1-4}1 &1.8 $\pm$ 0.2&1.03 $\pm$ 0.01&14.2 $\pm$0.3 \\

\cline{1-4} 2 & 1.76 $\pm$ 0.05&2.04 $\pm$ 0.03 &15.4 $\pm$ 0.3\\

\cline{1-4} 3 & 1.67 $\pm$ 0.02&3.01 $\pm$ 0.05 &17.6 $\pm$ 0.5\\

\cline{1-4} 4 & 1.68 $\pm$ 0.01& 3.97 $\pm$ 0.09 &20.1 $\pm$ 0.6\\

\cline{1-4} 6 & 1.66 $\pm$ 0.01& 5.99 $\pm$ 0.17 &22.0 $\pm$ 0.7\\

\cline{1-4} 8 & 1.66 $\pm$ 0.01& 8.13 $\pm$ 0.30 &26.0 $\pm$ 1.0\\

\cline{1-4}

\end{tabular}
\caption{\small{Reconstructed values of the single base-track resolution $\delta \theta_s$, the average momentum
 $\langle p \rangle$, and the momentum resolution $\frac{\Delta(1/p)}{(1/p)}$ for MC samples of tracks
crossing an entire brick and for different energies
 simulated with $\delta\theta_s^{MC}$ = 1.67 mrad.}}
\label{tab:dtMC}
\end{center}
\end{table}
\end{linenomath}

\par The values of  $\langle p \rangle$ and $\frac{\Delta(1/p)}{(1/p)}$ have been obtained with the 
method described in Section \ref{sec:explain}. 
It appears, that the linearity between reconstructed and MC momenta is kept over the whole range, 
 and that the momentum resolution worsens with the momentum, as expected.
The linearity of the MC reconstructed momentum and the evolution of
the momentum resolution show the consistency of the method. They also demonstrate that  the approximation of lead as the main scattering element
 is well-suited for the OPERA ECC configuration. 
\par
These results were obtained for tracks passing through 56 cells of an ECC brick. 
Figure~\ref{fig:resmom} shows how the resolution worsens with increasing momentum  and with decreasing track span.
\begin{linenomath}
\begin{figure}[!h]
\begin{center}
\includegraphics[width=95mm]{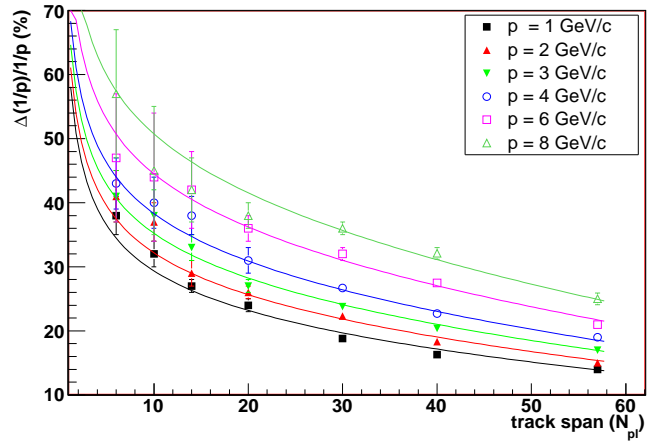}
\vspace*{-0.2cm} \caption{\small{Momentum resolution dependence on
track span $N_{pl}$ for MC pions with $\delta \theta_s^{MC}$ = 1.67 mrad. The solid lines correspond to the 
fitted parameterised resolution function of Eq.~\ref{mcseq15}.}}
\label{fig:resmom}
\end{center}
\end{figure}
\end{linenomath}
\par Using all the MC results for  different track spans and momenta values,
the momentum resolution $\frac{\Delta(1/p)}{(1/p)}$ has been parameterised in terms of the momentum $p$ and track span $N_{pl}$ as: 

\begin{linenomath}
\begin{equation}\label{mcseq15}
\frac{\Delta
(1/p)}{(1/p)}=(0.397+0.019\times p)/\sqrt{N_{pl}}+ (0.176+0.042\times  p)+(-0.014-0.003\times p)\times \sqrt{N_{pl}}.
\end{equation}
\end{linenomath}
The fitted function 
well describes  all momentum measurements from 1 to 8 GeV/c for various track lengths.

\subsection{Tracks at large incident angles} \label{sec:analm}

A first sample of 2, 4, and 6 GeV/c MC pions has been generated at  angles of $\theta_X$ = 200 and 400 mrad and
 $\theta_Y$ = 0 mrad. 
The $\delta\theta_T$ and $\delta\theta_L$ angular dependences have been parameterised according to Equations~\ref{eq:theta1} 
and \ref{eq:theta2} 
using the resolution parameters measured with a special brick consisting of a sequence of emulsion films,
without lead exposed to 7 GeV/c pions at several incident
angles. The track resolution parameters are measured to be 
$\delta\theta_L $(0) = $\delta\theta_T$(0) = 2.1 mrad, and $\epsilon_z$ = 9.3.

The different MC samples have been simulated using this parameterisation of the angular resolution. 
For the MC event samples at $\theta_X$ = 200 mrad, the measured values of $\langle p \rangle$ in 3D and in 2D projections T, L 
are compatible with the expected values.
The values of $\langle p \rangle$ in the 3D and L projections  for 4 GeV/c and 6 GeV/c pions at $\theta_X$ = 400 mrad 
are respectively 10\% and  20\% lower  than the true momentum while there is agreement in the T projection.
 This is explained by the angular dependence of the longitudinal resolution which increases linearly with the angle
reaching already a factor of two for track angles of 200 mrad.
Note, that the T projection is not affected since it is angle-independent.

For the same reason,
 the momentum resolution is stable, as can be observed in Figure \ref{fig:MCdepang} depicting the dependence 
of $\frac{\Delta(1/p)}{(1/p)}$ on the angle in the T projection. This plot also shows that the momentum
 resolution in the T projection is worse than in the 3D case for tracks at 0 mrad, due to the 50\% reduced statistics 
when using only one projection.
\begin{linenomath}
\begin{figure}[!h]
\begin{center}
\includegraphics[width=80mm]{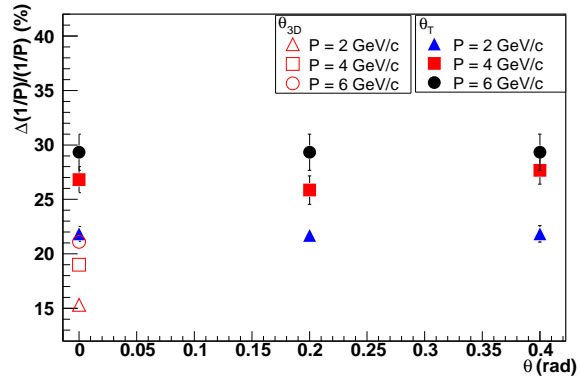}
\vspace*{-0.2cm} \caption{\small{Momentum resolution with respect to the 3D angle $\theta$ for different MC pion momenta,
 obtained using only the T projection. 
As a reference, the values obtained at 0 mrad in 3D are indicated by the open symbols.
}} \label{fig:MCdepang}
\end{center}
\end{figure}
\end{linenomath}

 All the previous considerations lead to the conclusion that at large angles, the optimal method for estimating the momentum is 
to use the T projection, which is not biased and not angle-dependent. However, at small angles, the 3D calculation
 remains statistically more accurate,
resulting in improved momentum resolution. In the algorithm, the threshold for large angles is set to 200 mrad, such that the
angular resolution $\delta\theta$ is kept independent and always equal to the value for $\theta$ = 0.\\

\par A second sample of 2 and 4 GeV/c MC pions has been generated with $\theta_X$ = 400 mrad 
and $\theta_Y$ = 200 mrad. Since the 3D angle is above 200 mrad, 
we report only the results obtained with the T projection. The value of $\delta\theta_T$ is fixed at 2.1 mrad.
 The measured values of $\langle p \rangle$
 and $\frac{\Delta(1/p)}{(1/p)}$ are given 
in Table \ref{tab:res200-400}. The momentum measurements are compatible with the input values, and the momentum resolutions agree 
with previous estimates.\\
\begin{linenomath}
\begin{table}[!h]
\begin{center}
\begin{tabular}{|c|c|c|}
\cline{1-3} $p_{MC}$ (GeV/c) &  $\langle p \rangle$ (GeV/c)&$\frac{\Delta(1/p)}{(1/p)}(\%)$ \\
\cline{1-3} 2 & 1.9 $\pm$ 0.1 &22 $\pm$ 1\\
\cline{1-3} 4 & 3.9 $\pm$ 0.2 &26 $\pm$ 1\\
\cline{1-3}
\end{tabular}
\caption{\small{Results on $\langle p \rangle$ and $\frac{\Delta(1/p)}{(1/p)}$ with the T projection for 2 and 4 GeV/c pions, with
$\theta_X$ = 400 mrad and $\theta_Y$ = 200 mrad.}} \label{tab:res200-400}
\end{center}
\end{table}
\end{linenomath}

\par The method using the T projection for angles larger than 200 mrad is thus validated. 
Similarly to Equation~\ref{mcseq15} for small angles, 
it is now possible to parameterise this dependence at large angles
as well, using the T projection and a similar analytic formula. 
It gives:\begin{linenomath}
 \begin{equation}\label{eq14}
\frac{\Delta
(1/p)}{(1/p)}=(1.400-0.022\times p)/\sqrt{N_{pl}}+ (-0.040+0.051\times  p)+(0.003-0.004\times p)\times \sqrt{N_{pl}}.
\end{equation}\end{linenomath}

This single  function, shown in Figure \ref{fig:evol0ang},  describes  all the MC results  from 1 to 8 GeV/c
for various $p$ and $N_{pl}$ values. As in Equation~\ref{mcseq15}, it is used to assign the confidence level-ranges of single-track momentum measurement.
\begin{linenomath}
\begin{figure}[!h]
\begin{center}
\includegraphics[width=95mm]{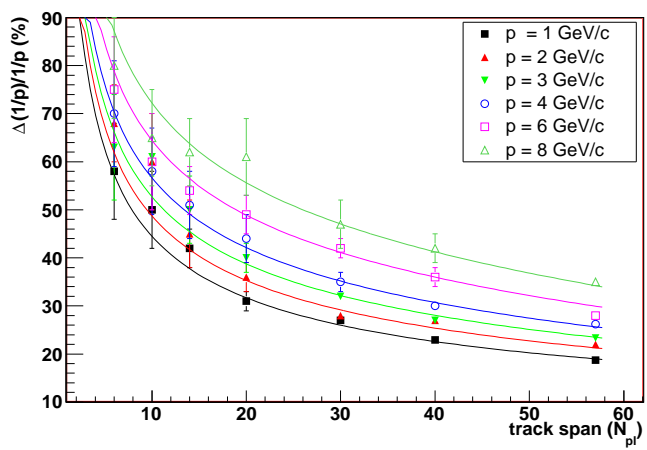}
\vspace*{-0.2cm} \caption{\small{Momentum resolution dependence on
track span $N_{pl}$ for MC pions using the T projection with $\delta \theta_s^{MC}$ = 1.67 mrad. The solid lines correspond to the 
fitted parameterised resolution function of Eq.~\ref{eq14}.}} \label{fig:evol0ang}
\end{center}
\end{figure}
\end{linenomath}

\subsection{Comments on method comparisons and systematics}
\par Various MC studies concerning systematic errors and comparisons with other methods have been carried out 
in~\cite{thesis}: 
\begin{itemize}
\item As explained in Section 2, using the differences between all combinations of pairs of angles separated by $N_{cell}$ 
cells increases the number of measurements.
It has been established that when using only the differences between successive pairs of angles, 
the fit of the momentum distributions diverges above 4 GeV/c. Moreover, the momentum resolutions are 1.5 times worse.
\item A track momentum $p$ can be measured  with both $\delta\theta$ (angular base-track resolution) 
and $p$ as free parameters in the fit procedure. 
However, an error of a few \% on $\delta\theta$ can affect the momentum reconstruction 
by more  than 10\% for high-energy tracks (above 4 GeV/c). 
   The best results are obtained with the proposed method keeping $\delta\theta$ fixed. The physical value of 
the base-track angular resolution is usually  between 1 and 2 mrad and depends mainly on the experimental conditions. 
The value of $\delta \theta$ can be determined or verified with reference measurements of angular 
deviations in  emulsion films, without scattering in heavy materials.
\item Effects from the correlations among the $\theta_{meas}$ values measured at the different $N_{cell}$ cells have been estimated
by building covariance matrices at  different energies and track lengths with MC samples. Fits of the track
 scattering angle dependence on $N_{cell}$ have been repeated by incorporating the covariance matrix in the minimization function
used to compute the track momentum. The difference with the uncorrelated fit is found to be less than a few percent for the
absolute momentum value determination for pion momenta ranging from 2 to 8 GeV/c and the resolution stays unchanged.

\end{itemize}

\section{Analysis of  pion test beam data} \label{sec:data}
We report here the results obtained with real data collected in a test beam-exposure of OPERA bricks to
2, 4, 6, and 8 GeV/c pions produced by the CERN PS accelerator. 
Figures \ref{fig:comp2g} and \ref{fig:comp6g} compare the momentum distributions of MC (red solid line) to real data (black crosses) for
pions crossing the entire brick with momenta of 2 and 6 GeV/c, respectively.
Table \ref{tab:data} summarises the values of the single base-track resolution $\delta \theta_s$, the average momentum
$\langle p \rangle$, and the resolution $\frac{\Delta(1/p)}{(1/p)}$ obtained for each data sample.
\begin{linenomath}
\begin{figure}[!h]
\begin{center}
\includegraphics[width=100mm]{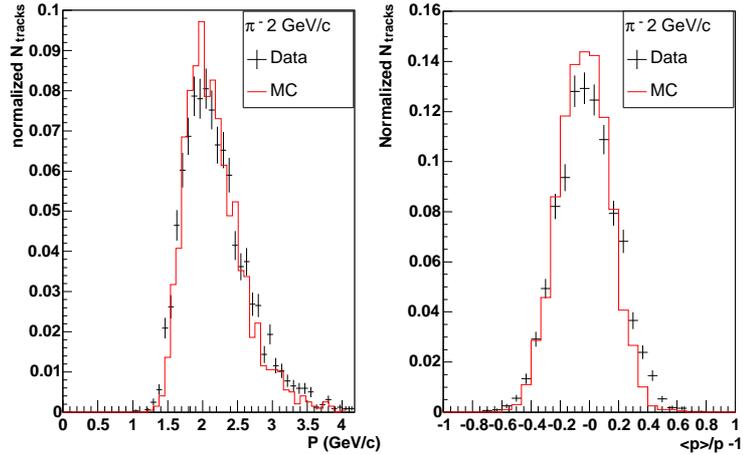}
\vspace*{-0.2cm} \caption{\small{Data/MC comparison for 2 GeV/c pions.
Left: momentum distribution. Right: inverted momentum
distribution  $(\langle p \rangle/p - 1)$.}} \label{fig:comp2g}
\end{center}
\end{figure}
\begin{figure}[!h]
\begin{center}
\includegraphics[width=100mm]{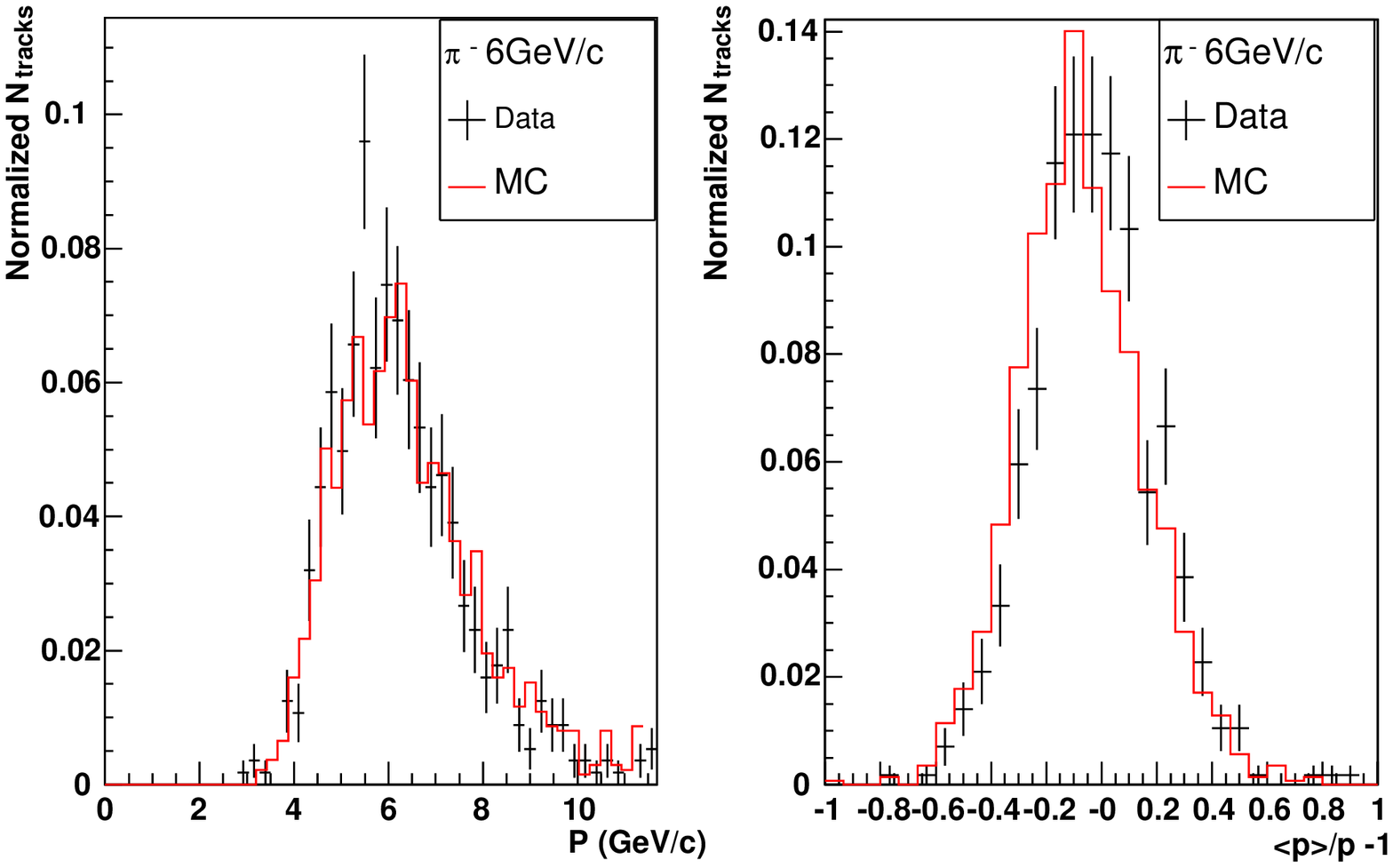}
\vspace*{-0.2cm} \caption{\small{Data/MC comparison for 6 GeV/c pions.
Left: momentum distribution. Right: inverted momentum
distribution $(\langle p \rangle/p - 1)$.}} \label{fig:comp6g}
\end{center}
\end{figure}
\end{linenomath}

\begin{linenomath}
\begin{table}[!h]
 \begin{center}
\begin{tabular}{|c|c|c|c|}
\cline{1-4} $p_{\pi} $ (GeV/c)&
 $\delta \theta_s$ (mrad)&
  $\langle p \rangle$(GeV/c)&
$\frac{\Delta(1/p)}{(1/p)}(\%)$ \\
\cline{1-4} 2 & 2.26 $\pm$ 0.01& 2.08 $\pm$ 0.05 & 19.6 $\pm$ 0.4\\
\cline{1-4} 4 & 1.72 $\pm$ 0.01& 4.32 $\pm$ 0.08 & 19.4 $\pm$ 0.4\\
\cline{1-4} 6 & 1.90 $\pm$ 0.01& 5.9 $\pm$ 0.2 & 21.0 $\pm$ 3.0\\
\cline{1-4} 8 &1.48 $\pm$ 0.01&7.2 $\pm$ 0.5 & 32.0 $\pm$ 2.0\\
\cline{1-4}
\end{tabular}
\caption{\small{Reconstructed values of $\delta \theta_s$,
$\langle p \rangle$, and $\frac{\Delta(1/p)}{(1/p)}$ obtained with  pion test beam data.}}
\label{tab:data}
\end{center}
\end{table}
\end{linenomath}

\normalsize{} The results for real and simulated data are consistent within  11\%.
Concerning additional systematic uncertainties coming from 
beam composition, it appears that while the 4 GeV/c and 6 GeV/c data sample have
 the expected $p$ resolution, the momentum resolution for 2 and 8 GeV/c
data is   measured to be respectively 4\% and 10\% worse than for the MC expectations. The discrepancy at 8 GeV/c
has been understood to come from a higher muon contamination produced after the
momentum selection collimators in the pion beam, which was not 
taken into account in the simulation. 
In the case of the 2 GeV/c sample, the  resolution is slightly worse due to scattering 
on  different materials placed along the beam line in front of the bricks.

Data at large angles from test beam pions of several energies and different incident angles recorded in one OPERA brick have
also been analysed. 
 Figure \ref{fig:slopesga} shows the angular distribution in $\theta_X$ 
for reconstructed tracks with length ($N_{pl}$) ranging from 25 to 30  plates. The different peaks correspond to: \begin{itemize}
\item 2 GeV/c pions at 200 and 400 mrad,
\item 4 GeV/c pions at -200 and -400 mrad,
\item 6 GeV/c pions at 100, 300 and 600 mrad,
\item 8 GeV/c pions at 50 mrad, used as reference data at small angles.
\end{itemize}

\begin{linenomath}
\begin{figure}[!h]
\begin{center}
\includegraphics[width=100mm]{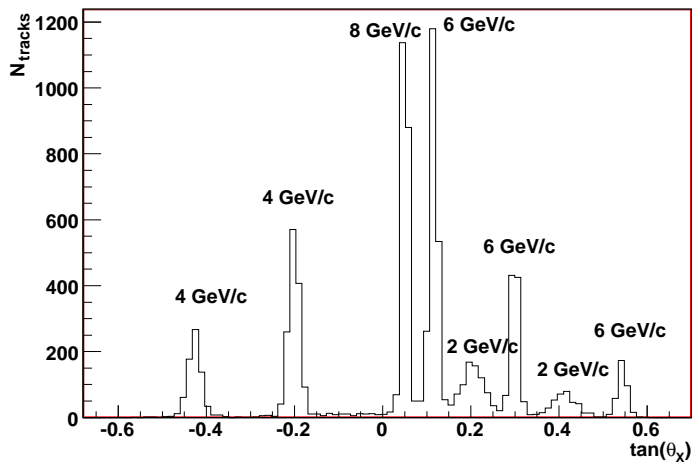}
\vspace*{-0.2cm} \caption{\small{Angular distribution in $\theta_X$ for pion tracks reconstructed in the brick with a span between 
25 and 30 plates.}} \label{fig:slopesga}
\end{center}
\end{figure}
\end{linenomath}

The results for large angles are summarised in Table~\ref{tab:data2}. The values of $\langle p \rangle$ are compatible with the expected 
pion beam momentum. The reconstructed momentum resolution can be compared to the one parameterised in
 Equation \ref{eq14}, also given in Table~\ref{tab:data2}. The measured values are compatible with the expectations except
 for the 2 GeV/c samples, where the measured value of $\frac{\Delta (1/p)}{(1/p)}$ is over-estimated by 25 to 40~\%. 
This effect is also due to the scattering on different materials, placed along the beam line in front of the brick during the test beam.
The 2 GeV/c pions have been particularly affected, as can be seen from the broad peaks in Figure \ref{fig:slopesga}: the 
interactions on materials lead to a dispersion in angle and in energy, which deteriorates the results on momentum resolution
 at low energies.\\

\begin{linenomath}
\begin{table}[!h]
 \begin{center}
\begin{tabular}{|c|c|c|c|c|c|}
\cline{1-6} $p_{true}$ (GeV/c) & $N_{pl}$ & $\theta_{3D} (rad)$&
  $\langle p \rangle$(GeV/c)&
$\frac{\Delta(1/p)}{(1/p)}$ &$\frac{\Delta(1/p)}{(1/p)}_{expected}$ \\
\cline{1-6} 2 &36 & 0.2 & 2.2 $\pm$ 0.2 & 37 $\pm$ 5\%&26\%\\
\cline{1-6} 2 &28 & 0.4 & 2.1 $\pm$ 0.1 & 38 $\pm$ 3\%&30\%\\
\cline{1-6}\multicolumn{6}{|c|}{}\\
\cline{1-6} 4 &36 & 0.2 &4.3 $\pm$ 0.3 & 32 $\pm$ 2\%&32\%\\
\cline{1-6} 4 &28 & 0.4 &4.0 $\pm$ 0.5& 42 $\pm$ 6\%&37\%\\
\cline{1-6}\multicolumn{6}{|c|}{}\\
\cline{1-6} 6 &36 & 0.1 &6.3 $\pm$ 0.6 & 44 $\pm$ 5\%&38\%\\
\cline{1-6} 6 &36 & 0.3 & 6.1 $\pm$ 0.6 & 38 $\pm$ 4\%&38\%\\
\cline{1-6} 6 &28 & 0.5 & 5.7 $\pm$ 0.5 & 45 $\pm$ 4\%&44\%\\
\cline{1-6}
\end{tabular}
\caption{\small{Results of momentum measurements obtained with the pion test beam at different angles and energies.
 The calculation was done in the T projection, with $\delta\theta_T$ fixed at 2.1 mrad.}} \label{tab:data2}
\end{center}
\end{table}
\end{linenomath}
\normalsize{}

\par Taking into account these effects, one can conclude that MC and test beam data are compatible and give 
consistent results at both small and large angles.

\section{Soft muon momentum measurement in OPERA}
In order to validate the algorithm with charged particles produced in neutrino interactions, a sample of muons
recorded in the 2008 run originating from charged current interactions ($\nu_{\mu}^{CC}$) in the OPERA target bricks 
was selected.
Details on the detector characteristics, data acquisition, event reconstruction, and analysis procedures are described in~\cite{opera_run}.
 More details on the performance of the 
OPERA electronic detectors can be found in~\cite{ED}.

The muon momentum in the electronic detectors was obtained either from the range of the particle 
in the OPERA target tracker, or in the spectrometer yoke, or from the magnetic spectrometer measurement. The   
corresponding momentum resolution $\Delta p/p$ is estimated at about 10\% for the analysed sample.
In order to match the momentum range
accessible with the MCS algorithm, charged current interactions where  a  muon was reconstructed in the electronic 
detectors with a momentum below 6 GeV/c were selected. 
The corresponding neutrino interaction 
vertices were located in the emulsion target, and one emulsion track per event was matched to the muon track predicted by the
 electronic detectors.
Additional selection criteria have been applied on track quality and length. 
The required minimum track span  is 10 cells. The final sample
corresponds to 43 events. Figure~\ref{fig:softmutrack} shows the dependence of the angular deviation
on the thickness  of lead traversed in the 3D (left) and the T (right) coordinates for two different muon tracks. 
\begin{linenomath}
\begin{figure}[!h]
\begin{center}
\includegraphics[width=70mm]{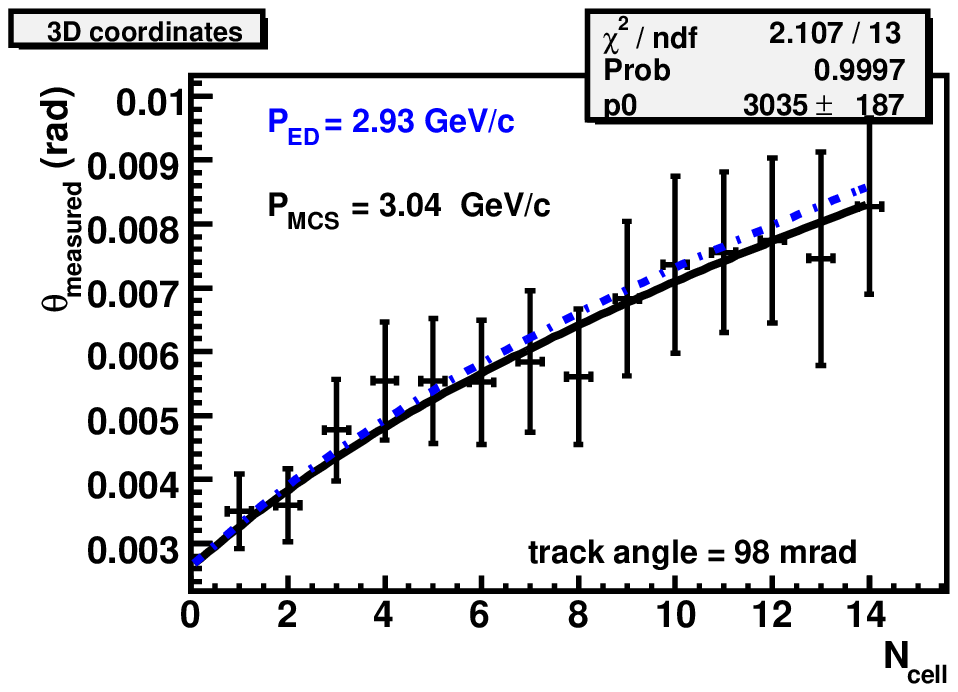}
\includegraphics[width=70mm]{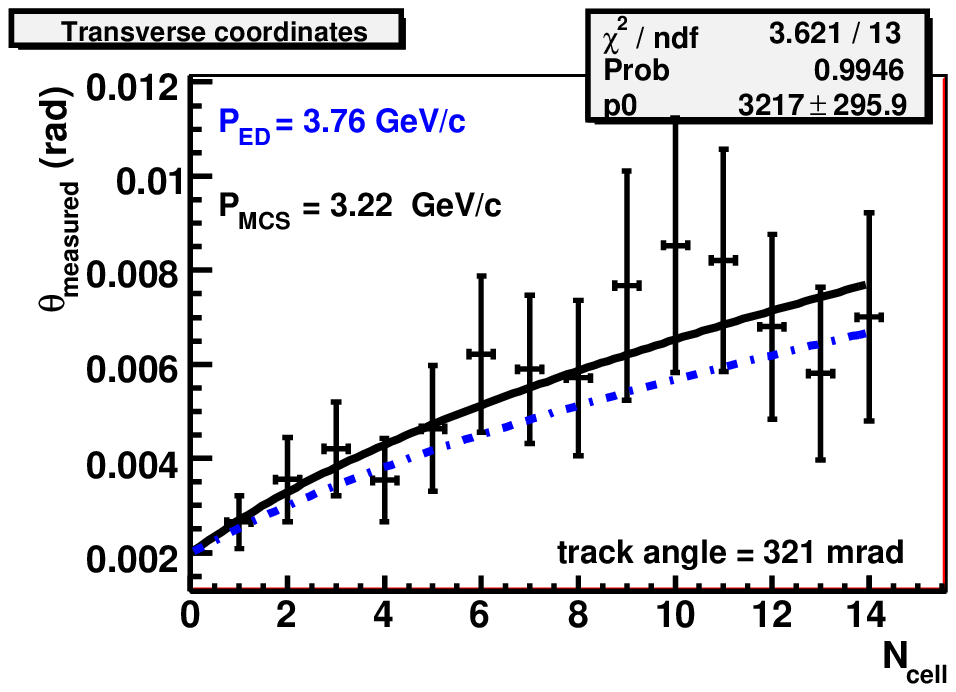}
\vspace*{-0.2cm} \caption{\small{Angular deviation dependence on the thickness of the lead traversed by a muon track
with an incident angle of 98 mrad using the 3D coordinates (left) and for a muon track with an incident 
angle of 321 mrad using the T coordinate (right).
The dashed line shows the expected angular dependence obtained 
with the momentum measured
by the electronic detectors while the solid line corresponds to the momentum measured by the MCS algorithm in emulsion.}} \label{fig:softmutrack}
\end{center}
\end{figure}
\end{linenomath}
The dashed line shows the expected angular dependence obtained 
with the momentum measured
by the electronic detectors while the solid line is the result of the fit of the momentum by the MCS
method described in this paper. The two momentum measurements from MCS and electronic detectors  are fully compatible.

The muon momenta in the selected sample range from 2 to 6 GeV/c, as
can be seen in the left plot of 
Figure~\ref{fig:mcs1}, which shows the correlation between the two measurements:  the right plot shows  the relative difference
with respect to the electronic detector value.
\begin{linenomath}
\begin{figure}[!h]
\begin{center}
\includegraphics[width=60mm]{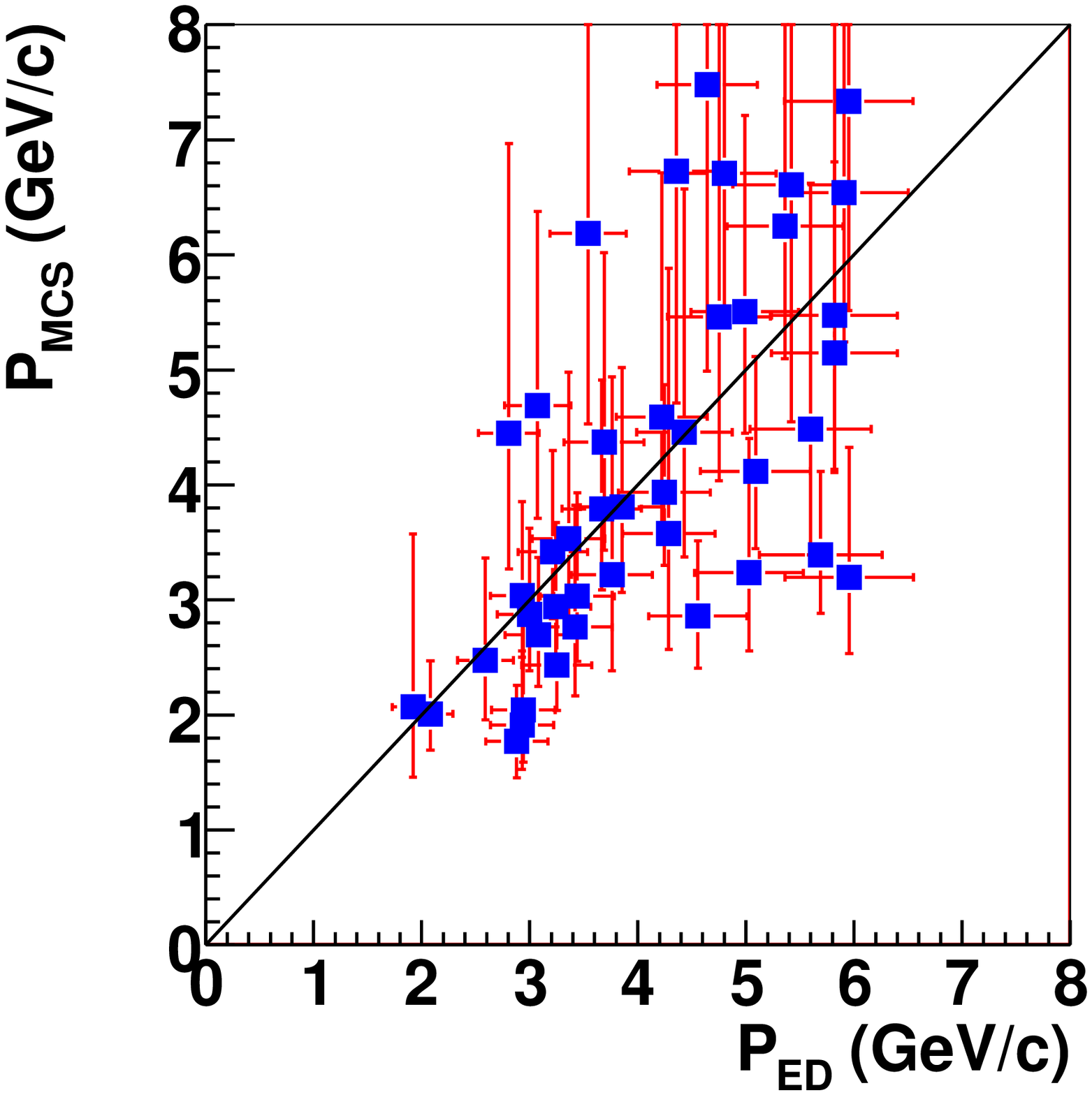}
\includegraphics[width=60mm]{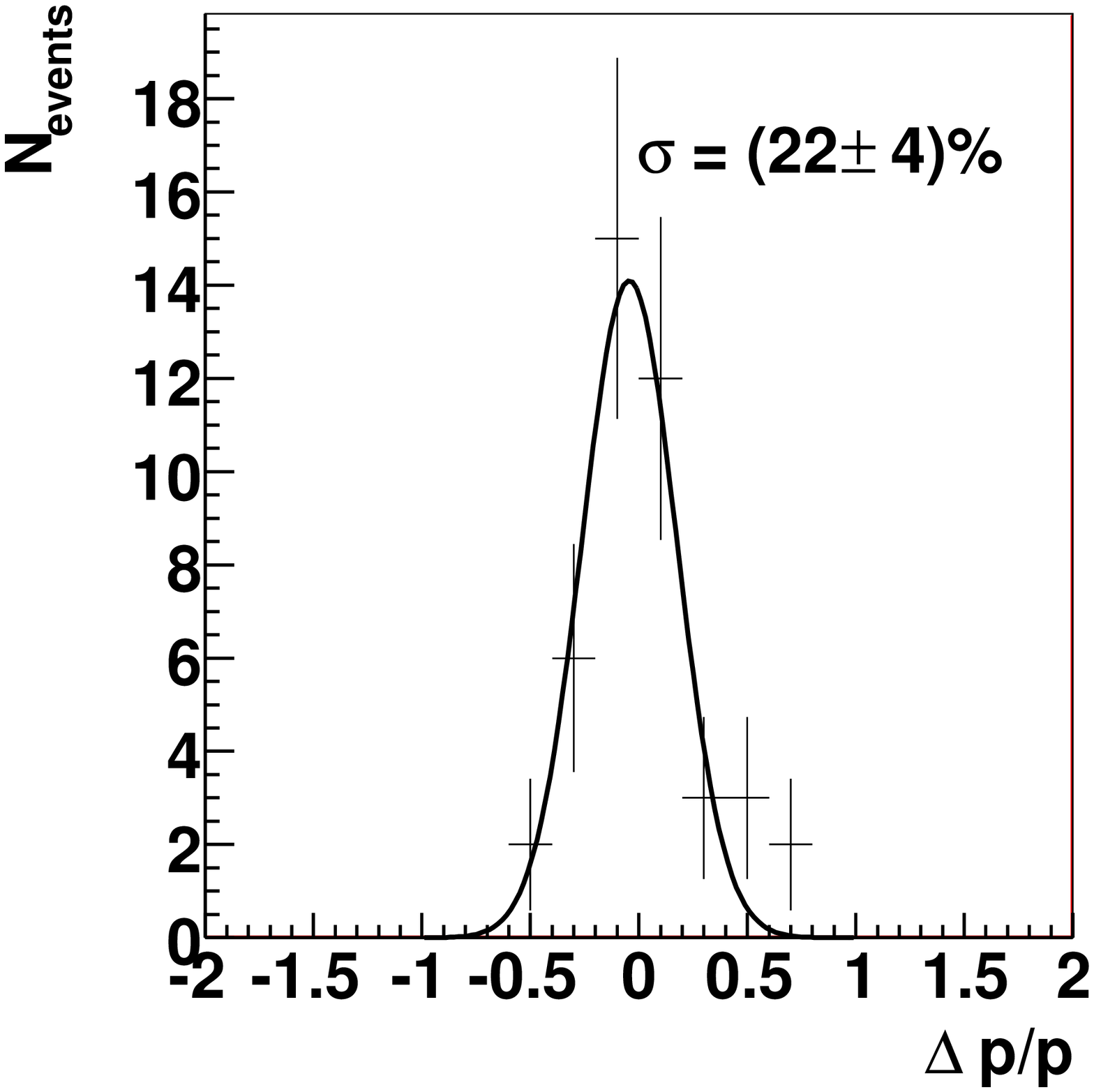}
\vspace*{-0.2cm} \caption{\small{Left: Muon momenta measured by MCS (P$_{MCS}$) as a function of the momenta
 obtained from the electronic detectors (P$_{ED}$). The error bars 
correspond to the 68\% confidence level range. Right:  The relative difference
between the two measurements with respect to the electronic detector measurement. }} \label{fig:mcs1}
\end{center}
\end{figure}
\end{linenomath}

The distribution is a Gaussian centered at zero. The width gives an average resolution of $(22\pm 4)$\%, compatible with the 
expectation obtained by folding the track sample characteristics with the  parameterised resolution functions. The width includes
also a contribution from the electronic detector resolution. 
In order to cross-check the  estimate of the experimental uncertainty, the differences of the measured inverted momenta have been normalised to 
the uncertainty estimates on $1/p$, given by  Equations~\ref{mcseq15} and \ref{eq14} for the different track spans and angles.
The resulting Gaussian distribution has a  standard deviation of $1.10\pm0.24$, compatible with unity.
This shows that the uncertainty for each track is properly estimated.

\section{Conclusions}

An improved angular method has been developed to exploit Multiple Coulomb Scattering for the momentum measurement of 
charged particles in Emulsion Cloud Chamber detectors. 
The results of Monte Carlo  studies and pion test beam data show that  momenta
 up to 8 GeV/c can be measured with a resolution better than 30$\%$.
The approach has been optimized for small incident angles, as well as for large-angle 
tracks  entering the OPERA lead-emulsion target elements and is well-suited for the neutrino interaction analysis. 
The results obtained with muons measured with the OPERA electronic detectors have confirmed the validity of the approach and 
assessed the performance of the  algorithm.

\section{Acknowledgements}

We thank CERN for the commissioning of the CNGS facility and for its
successful operation, we thank INFN for the continuous support given to the
experiment during the construction, installation and commissioning phases
through its LNGS laboratory. We warmly acknowledge funding from our
national agencies: Fonds de la Recherche Scientifique - FNRS and Institut
Interuniversitaire des Sciences Nucl{\'e}aires for Belgium, MoSES for 
Croatia, CNRS and IN2P3 for France, BMBF for Germany, INFN for Italy,
JSPS (Japan Society for the Promotion of Science), MEXT (Ministry of
Education, Culture, Sports, Science and Technology ), QFPU (Global COE 
program of Nagoya University, ``Quest for Fundamental Principles in the
Universe" supported by JSPS and MEXT) and Promotion and Mutual Aid
Corporation for Private Schools of Japan, 
SNF, the Canton of Bern and the ETH Zurich for Switzerland,
the Russian Foundation for Basic Research (grant 09-02-00300 a), 
Programs of the Presidium of the Russian Academy of Sciences ``Neutrino Physic'' and
``Experimental and theoretical researches of fundamental interactions connected with 
work on the accelerator of CERN'', Programs of support of leading schools (grant 3517.2010.2),
and the Ministry of Education and Science of  the Russian Federation for Russia,
the Korea Research Foundation Grant (KRF-2008-313-C00201) for Korea and
TUBITAK, The Scientific and Technological Research Council of Turkey, for Turkey. 
We are also indebted to INFN for providing fellowships and grants
to non Italian researchers. We thank the IN2P3 Computing Centre (CC-IN2P3)
for providing computing resources for the analysis and hosting the central
database for the OPERA experiment. We are indebted to our technical
collaborators for the excellent quality of their work over many years of
design, prototyping and construction of the detector and of its facilities.
Finally, we thank our industrial partners.

\end{document}